\documentclass[a4paper,11pt]{article}
\usepackage{pos}

\usepackage{natbib}
\title{The spectral reconstruction problem for thermal photon and dilepton rates }

\author*[a]{Anthony Francis}

\affiliation[a]{National Yang Ming Chiao Tung University,\\
  1001 Daxue Road, Hsinchu, Taiwan}

\emailAdd{afrancis@nycu.edu.tw}

\abstract{Thermal photon and dilepton rates are important probes for understanding the quark-gluon plasma and QCD at high temperatures.
As a consequence there is a strong interest to determine them using lattice QCD calculations. However, this is made difficult as they are related to thermal spectral functions that are not directly accessible through lattice calculations. Instead they are indirectly obtainable through performing an inverse Laplace-type transformation of Euclidean time lattice correlation functions. In this talk we will present recent results in dynamical QCD with a focus on advancements in determining the photon production rate via spectral reconstruction from lattice data.
}

\FullConference{The XVIth Quark Confinement and the Hadron Spectrum Conference (QCHSC24)\\
 19-24 August, 2024\\
 Cairns Convention Centre, Cairns, Queensland, Australia\\}

\begin{document}
\maketitle

\section{Motivation}

At this conference many of the contributions that are addressing different QCD physics topics are facing a similar core issue in the form of an inverse problem. Indeed, inverse problems are common across science in general. The common premise is that there is a mismatch between the available and the desired information. Given this mismatch we need a robust map, or transformation, from one into the other, but in the case of an inverse problem this transformation is (numerically) ill-posed or ill-conditioned.\\
Focusing on particle physics observables, and in particular their theoretical calculation from Euclidean quantum field theory using lattice QCD calculations, some examples of quantities that face an inverse problem are: $N \to N'$ scattering amplitudes at any $s = E^2_{\sf cm}$, 
$N + j \to N'$ transitions at any $s$, non-local matrix elements like the $R$-ratio or the hadronic tensor \cite{Hansen:2017mnd,ExtendedTwistedMass:2024myu}, inclusive processes for deep inelastic scattering \cite{Fukaya:2020wpp} and semileptonic decays \cite{Gambino:2020crt}, and also the determination of distribution functions (PDFs, GPDs, TMDs) \cite{HadStruc:2024rix}. These are just a few quantities that can be extracted from a Euclidean expectation value by solving an inverse problem at $T=0$. At finite temperature, $T>0$, solving an inverse problem gives access to transport coefficients, production rates and thermal broadening, for example bulk and shear viscosities \cite{Meyer:2007dy,Meyer:2007ic,Altenkort:2022yhb}, the electrical conductivity \cite{Aarts:2007wj,Ding:2010ga,Amato:2013naa,Aarts:2014nba,Aarts:2020dda} and heavy quark diffusion \cite{Caron-Huot:2009ncn,Banerjee:2011ra,Francis:2015daa,Altenkort:2020fgs,Altenkort:2023oms} in the quark-gluon plasma (QGP), and also the dilepton and photon production rates \cite{Ding:2010ga,Ghiglieri:2016tvj,Brandt:2017vgl}, see \cite{Kaczmarek:2022ffn} for a recent overview and further recent references below.
Robust procedures and methods to perform spectral reconstructions, i.e. solve an inverse problem, or to find ways to circumvent them are therefore of broad interest and can impact a variety of fields.

\section{Probes of the quark gluon plasma}

\begin{figure}
\centering
\includegraphics[width=0.33\textwidth]{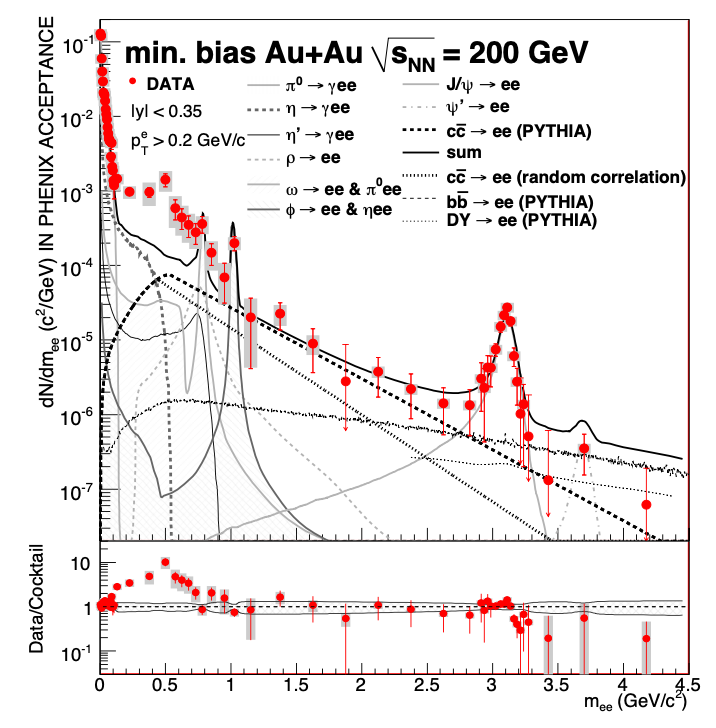}
\hspace{3ex}
\includegraphics[width=0.33\textwidth]{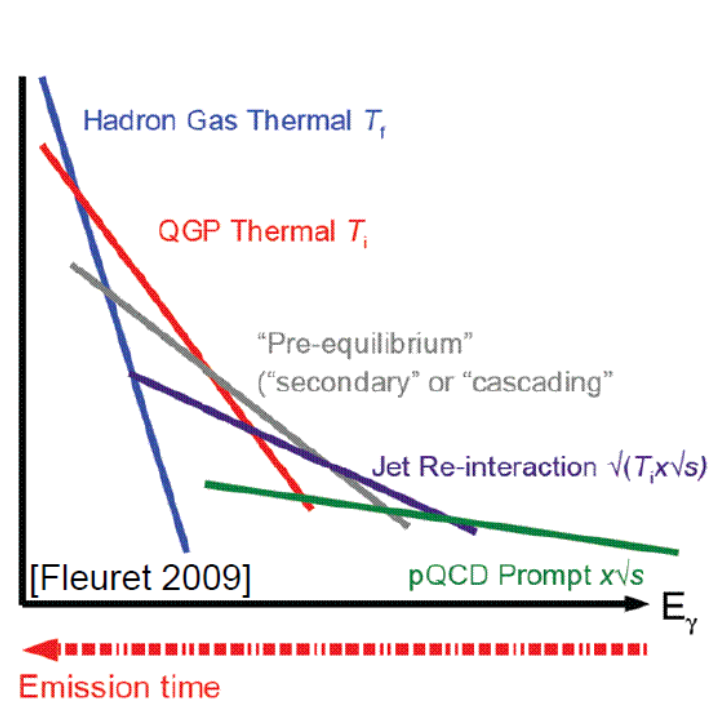}
\caption{ Left: Example of the thermal dilepton production rate determined in experiment, PHENIX \cite{PHENIX:2009gyd}. Right: Sketch of different sources of photons and their dominant energy regimes from \cite{Fleuret:2009zza}.}
\label{fig:dileptons-ex}
\end{figure}

In this contribution we focus on dilepton and photon production rates, but aim to also make the embedding of their study in the broader landscape of inverse problems clear.
Dilepton and photon production rates are especially interesting to study because lepton pairs and photons are produced at every stage of a heavy-ion collision (HIC). Once they are formed they have only a weak coupling to the QCD degrees of freedom that make up the QGP. As a consequence they effectively decouple after production and can be detected with little further modification. In this sense they probe the full thermal evolution of the HIC and the QGP medium. An example experimental result produced by PHENIX \cite{PHENIX:2009gyd} is shown in Fig.~\ref{fig:dileptons-ex} (left) alongside a sketch \cite{Fleuret:2009zza} of the regimes where the different thermal stages dominate the signal (right). 

Our goal is to understand the excess in the regime around $m_{ee}\simeq 0.5~$GeV in dilepton production and the dominant regime for $p_T\simeq (1-2)~$GeV in photon production from theory. To this extent we first note that the photon emissivity is related to thermal vector-vector current spectral functions (notation of \cite{MeyerCERN24}):
\begin{equation}
\rho^{\mu \nu}(\mathcal{K})=\int d^4 x e^{i \mathcal{K} \cdot x} \frac{1}{Z} \sum_n e^{-E_n / T}\left\langle n\left|\left[j^\mu(x), j^\nu(0)\right]\right| n\right\rangle
\label{eq:photonrate}
\end{equation}
Within these, the production rates that we want are the special cases 
\begin{align}
&\text{Dilepton rate: } \quad d \Gamma_{\ell^{+} \ell^{-}}(\mathcal{K}) =\alpha^2 \frac{d^4 \mathcal{K}}{6 \pi^3 \mathcal{K}^2} \frac{-\rho^\mu{ }_\mu(\mathcal{K})}{e^{\beta \mathcal{K}^0}-1} \quad\text{where  ~~}\mathcal{K}^2 \equiv \omega^2-k^2\\
&\text{Photon rate: } \quad  d \Gamma_\gamma(\boldsymbol{k})=\alpha \frac{d^3 k}{4 \pi^2 k} \frac{-\rho^\mu{ }_\mu(k, \boldsymbol{k})}{e^{\beta k}-1}\\
&\text{Electrical conductivity: } \quad  \sigma_{e l}=e^2 \sum_{f=1}^{N_f} Q_f^2 \lim _{k \rightarrow 0^{+}}  \frac{\rho^{i}_{~i}(k, \boldsymbol{0})}{k}
\end{align}
and we see that, apart from known factors, they are given by different regimes in these spectral functions:
\begin{align}
&\text{Dilepton rate: }\quad \Gamma_{\ell^{+} \ell^{-}}(\mathcal{K})\sim-\rho^{\mu}_{~\mu}(\mathcal{K})\rightsquigarrow \rho^{\mu}_{~\mu}(\omega,\boldsymbol{k}=0)\\
&\text{Photon rate: }\quad \Gamma_{\gamma}(\boldsymbol{k})\sim-\rho^{\mu}_{~\mu}(k,\vec k)\rightsquigarrow \rho^{\mu}_{~\mu}(\omega=k,\boldsymbol{k}\neq 0)\\
&\text{Electrical conductivity: }\quad\sigma_{e l}\sim \rho^{i}_{~i}(0, \boldsymbol{0}),~~\quad\rho^{0}_{~0}(0, \boldsymbol{0})=const.=:\chi_q
\end{align}

For the dilepton and photon rates we are particularly interested in the low-energy, non-perturbative, regime of QCD. A careful comparison and combination of perturbative and non-perturbative calculations, typically through lattice QCD, is then one of the goals.
Hereby lattice calculations at $T>0$ work in the imaginary-time path-integral representation of quantum field theory, i.e. the Matsubara formalism. As a consequence only imaginary-time vector correlators are accessible on the lattice ($\tau=it$)
\begin{equation}
	G^{\mu \nu}\left(\tau, \boldsymbol{k}\right)=\int d^3 x e^{-i \boldsymbol{k} \cdot \boldsymbol{x}}\left\langle j^\mu(x) j^\nu(0) \right\rangle_T=\int d^3 x e^{-i \boldsymbol{k} \cdot \boldsymbol{x}} \operatorname{Tr}\left\{\frac{e^{-\beta H}}{Z(\beta)} j^\mu(x) j^\nu(0)\right\}
\end{equation}
where $x=(\tau,\vec x)$, $j^\mu=\sum_f Q_f \bar{\psi}_f \gamma^\mu \psi_f$ is the vector current, $\psi$ are the involved quark fields, $f$ is the flavor index, $Q_F$ the electric charge and $\gamma^\mu$ acts on Dirac space. The factor $Z(\beta)$ is the QCD partition function, $\beta=1/T$ and $H$ the QCD Hamiltonian.
Their spectral representation is: 
\begin{equation}
G^{\mu \nu}\left(\tau, \boldsymbol{k}\right)=\int_0^{\infty} \frac{d \omega}{2 \pi} ~{\rho^{\mu \nu}(\omega, \boldsymbol{k})} ~\frac{\cosh \left[\omega\left(\beta / 2-\tau\right)\right]}{\sinh (\beta \omega / 2)} ~~.
\end{equation}
From such a Euclidean correlator calculated in lattice QCD (on the left) we want to extract the spectral function (on the right). As such we want to perform a simultaneous Wick rotation and Fourier transform. Note that formally
\begin{equation}
\rho(\omega)=\mathcal{L}^{-1}\left\{G_E(\tau)\right\}=\frac{1}{2 \pi i} \int_{\gamma-i \infty}^{\gamma+i \infty} e^{\omega \tau} G_E(\tau) d \tau\quad \textrm{and} \quad \rho(\omega)=\frac{1}{\pi}\textrm{Im}\left(G_M(-\omega)\right)
\end{equation}
which gives indication that the problem is related to having only real data where complex information is required.

\section{Properties of the inverse problem}

Not all inverse problems are created equal and some are more readily approachable than others. This has to do with the type of kernel that enters and sets the specific type of problem:
 \begin{equation*}
     G_{\rm{E}}(\tau) = \int_0^\infty \! \! d \omega \, \rho(\omega) \, \kappa(\omega, \tau) \,.
 \end{equation*}
 At zero temperature the kernel is $\kappa(\omega, \tau) = e^{- \omega \tau}$ and therefore the task is to perform an inverse Laplace transform. For distributions functions, e.g. the quasi-PDFs \cite{Karpie:2019eiq} it is $\kappa(\nu, x) = \cos(\nu x) \, \Theta(1 - x)$ and the inverse problem becomes a Fourier transform. In our case, at finite temperature, $\kappa(\omega, \tau) = \frac{\cosh(\omega(\beta/2 - \tau))}{\sinh(\omega \beta/2)}$, which turns into the inverse Laplace transform for $\lim_{T\rightarrow 0}$. However, in lattice calculations $T=1/N_\tau$ 
 and temporal extents are so short that a strong $\cosh$-behavior is visible. This could indicate a benefit for calculations on anisotropic lattices. Further, the shape of the kernel suppresses low-$\omega$ contributions at short $\tau$, while these low-$\omega$ contributions compete with the kernel $T$-effects at $\tau=N_\tau/2$.  For the quasi-PDFs some headway has been made \cite{Karpie:2019eiq} and new methods are being applied with success at $T=0$ \cite{Hansen:2017mnd,ExtendedTwistedMassCollaborationETMC:2022sta,ExtendedTwistedMass:2024myu}, however, the $T>0$ situation remains unique in the challenges it poses also due to the kernel.

Recalling the initial formulation by Jacques Hadamard a well-posed problem \cite{hadamard} should satisfy the three conditions of, the existence of a solution (1), the uniqueness thereof (2) and that the solution is stable in the sense that it changes continuously with the initial conditions (3). The inverse problem considered here fails these conditions in the sense of (3) due to our limitation to just a discrete sampling of the left hand side plus a finite precision. As such this is a method-independent statement. 
As a further side remark, more careful study reveals that our problem is ill-conditioned in principle \cite{Cuniberti:2001hm}. The proof provided shows that a solution can in principle be constructed from a finite number of Laguerre polynomials. The problem then shifts to how many points at what precision are required so that sufficiently many terms can be constrained. Qualitatively it was argued that the precision would have to be drastically improved over the status quo \cite{Meyer:2011gj}. With often exponentially bad signal-to-noise ratios this is difficult to achieve.
\begin{figure}
\centering
\includegraphics[width=0.45\textwidth]{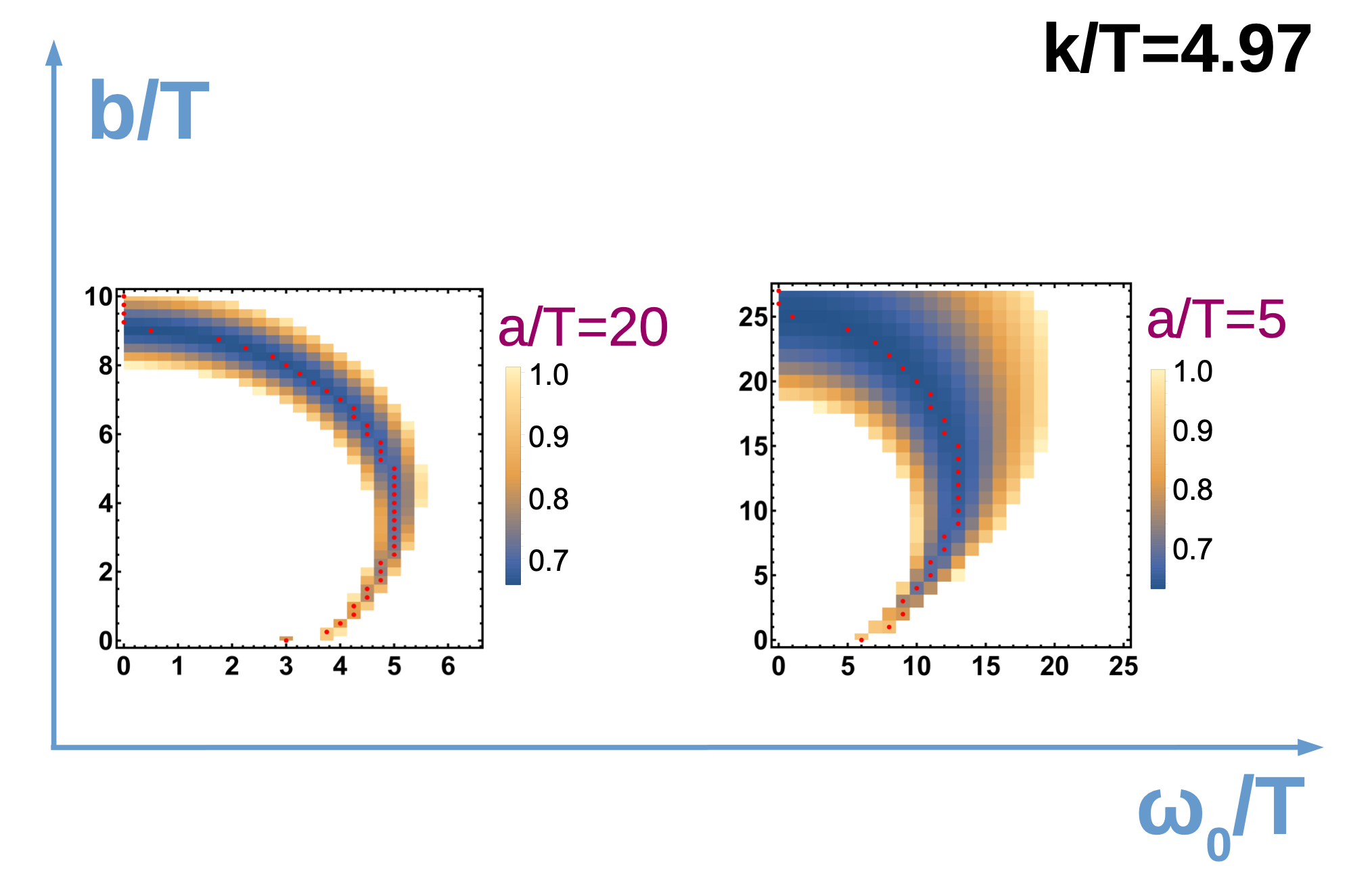}
\caption{ Example taken from \cite{Brandt:2017vgl}: Uncorrelated $\chi^2$-landscape over a three parameter Ansatz to solve the inverse problem of the photon rate via fit.
The example demonstrates the difficulty in finding stable solutions to the inverse problem that permit robust and precise statements.}
\label{fig:chi-landscape}
\end{figure}
To illustrate this, consider the following example where the spectral function $\rho(\omega)$ is determined via a $\chi^2$-fit \cite{Brandt:2017vgl}. Given lattice data $G_{data}(\tau,p)$ of the type to determine the photon production rate directly, one may motivate an Ansatz with 3 parameters $a, b,\omega_0$:
\begin{equation}
\chi_{f i t} = \left[G_{data}(\tau, p)-\int d \omega ~\rho_{f i t}\left(\omega, a,b,\omega_0\right) \frac{\cosh(\omega(\beta/2 - \tau))}{\sinh(\omega \beta/2)}\right] 
\end{equation}
In Fig.~\ref{fig:chi-landscape} the resulting $\chi^2$-landscape, i.e. the achieved $\chi_{fit}$ over the parameters is shown. We see that the problem takes the form of not being able to observe a clear global minimum. Each point on the map represents one "acceptable" set of parameters and therefore one solution that describes the Euclidean data.
Once they are somehow averaged or combined most features are washed out and robust statements can no longer be derived apart from the most broadest findings. In certain cases this can represent an advancement, however, in cases like the photon rate this is not enough. To shrink down these trenches in $\chi^2$ to wells or points one might ask if a brute force accuracy increase could lead to a more constrained result, but the point that we wanted to illustrate remains the same.

 \section{Approaches to the inverse problem}

The preceding example already demonstrates one approach to solving an inverse problem via $\chi^2$-fit. In general all methods to perform a spectral reconstruction can be understood as a master function
\begin{equation}
\mathcal F[\textbf G, \textbf C_{G}] = \big ( \, \pmb \rho, {\textbf C}_{\rho}  \, \big )
\end{equation}
 where $\textbf G$ represents a discrete set of samples $G(\tau)$, $\textbf C_{G}$ is the covariance of $\textbf G$, $\pmb \rho$ is a discrete estimator of $\rho(\omega)$ and  $\textbf C_{\rho}$ is the covariance of $\pmb \rho$. Our goal is to understand the properties and limitations of the master function $\mathcal F$. Hereby it is important to be focused on the data since the best $\mathcal F$ will depend in detail on $G$, number of slices, properties of $C$, and so forth.
 The general difficulty is that for $\pmb \rho_i = \rho(\omega_i)$ the quantity 
 \begin{equation}
    \big \vert  \mathcal F[\textbf G + \delta \textbf G, \textbf C_G + \delta \textbf C_G] -\mathcal F[\textbf G, \textbf C_G] \big  \vert \ \  \ \ \ \text{and thus} \ \  \ \ \ \   \big  \vert \textbf C_{\rho} \big  \vert  
    \end{equation}
 grow without bounds. At the same time, for cases where $\vert \textbf C_{\rho} \vert $ is under control, the relation between the target $\rho(\omega)$ and the discrete estimator $\pmb \rho$ may be obscured.
Faced with the premise that our goal is to reconstruct the spectral function encoded in the data. Generically one may follow the following distinct strategies:
\begin{itemize}
\item \textbf{Strategy I: Accept the premise and focus on information in the data.}\\
Examples of methods that use this strategy are approaches based on sparse modeling \cite{Itou:2020azb}, neural networks \cite{Kades:2019wtd} and recently Gaussian processes \cite{Horak:2021syv}. These are typically non-linear methods.
Challenges in this strategy are how to properly include and choose the information and constraints. This can be for example choosing the optimal bases for the sparse modeling or the correlation structures for Gaussian processes. Constraints that one might want to consider are the positivity of the spectral function and those posed by sum rules.
\item\textbf{Strategy II: Accept the premise and try to supply as much extra information as possible.}\\
Here examples are the mentioned $\chi^2$-fits, e.g. \cite{Ding:2010ga}, stochastic inference and optimization \cite{Ding:2017std} and also the variants of the Maximum Entropy Method (MEM) \cite{Asakawa:2000tr,Burnier:2013nla}. In the many cases the methods used are non-linear methods.
The challenges are how to effectively encode more specific information into e.g. the priors and to control any bias or systematics introduced.
\item\textbf{Strategy III: Reject the premise and focus on smeared spectral functions or new Euclidean quantities.}\\
Main examples of methods following this strategy are the Backus-Gilbert \cite{Backus:1968svk} and Hansen-Lupo-Tantalo \cite{Hansen:2019idp} methods, Gaussian processes and also expansion methods \cite{Gambino:2020crt}. These are often linear methods.
The challenge is to identify the physics observables that can benefit from smeared spectral function input or simplify or even avoid inverse problems altogether.
\end{itemize}
Each method has its own problem-dependent pros and cons. Following the argument that the issue can be fundamentally seen as related to the underlying data and a mismatch between it and the information we want, it is unlikely that there is a single best method or strategy to follow. Instead the best solution will likely cater to the specific problem at hand. One path that has been followed also in the following is to deploy approaches from all three strategies to the same data and to draw conclusions from their combined insights. 

\begin{figure}[t!]
\centering
\includegraphics[width=0.32\textwidth]{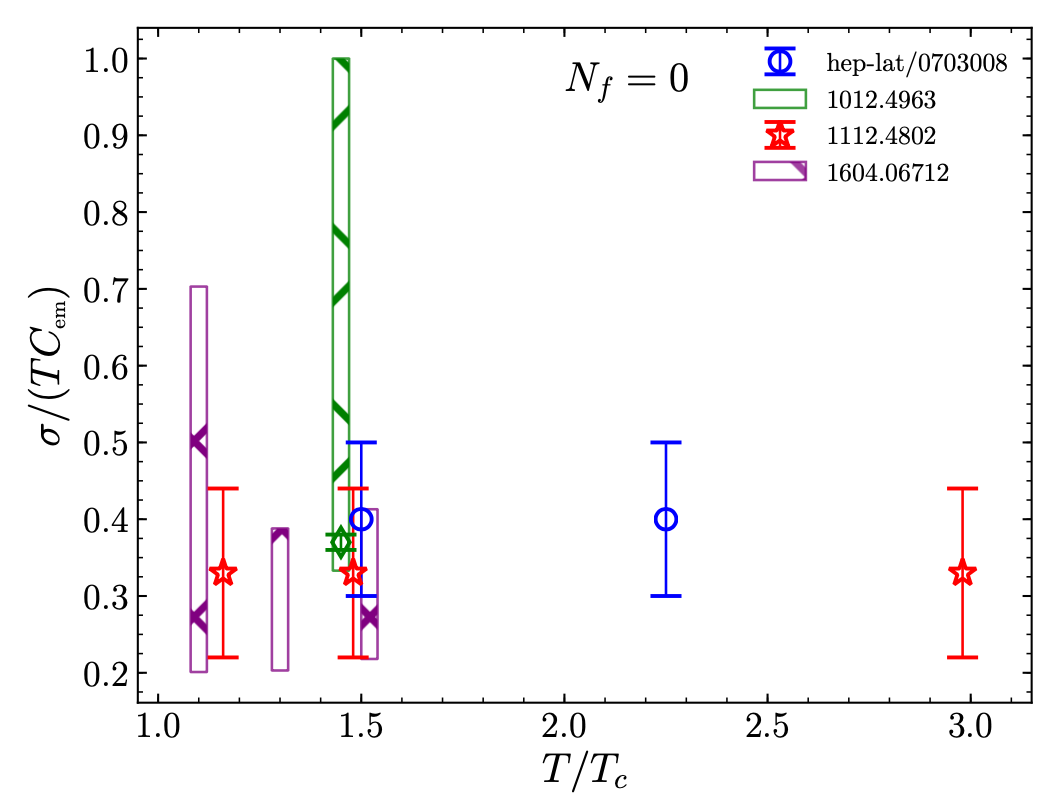}
\includegraphics[width=0.32\textwidth]{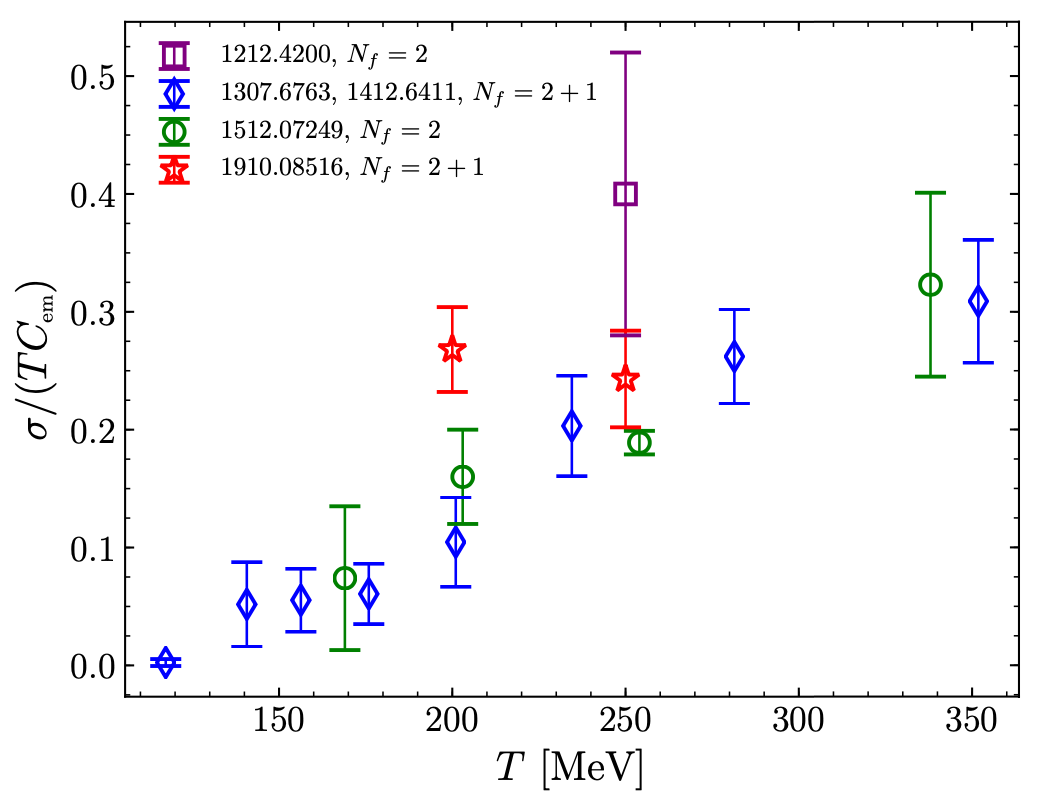}
\includegraphics[width=0.32\textwidth]{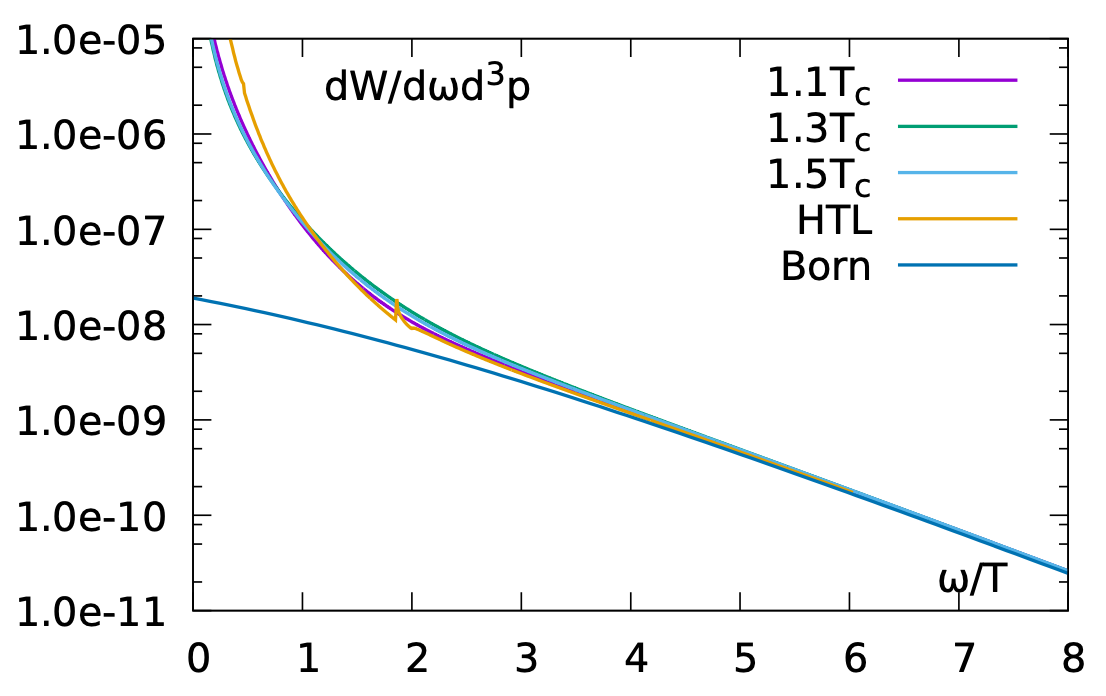}
\caption{Left and middle: Overview figures of the determinations of the electrical conductivity in quenched and dynamical QCD calculations, gathered in \cite{Aarts:2020dda}. Right: The related dilepton rate from \cite{Ding:2016hua}.}
\label{fig:comp-el}
\end{figure}

\section{Overview of results}

In the following a broad overview of the gathered results for the dilepton rates and the electrical conductivity is given. Afterwards we turn to the photon rate and show also new results \cite{Ali:2024xae}.
Generally speaking, lattice calculations of these spectral functions have to control effects due to statistical uncertainties (1), $\delta G(\tau)$ limits resolution; the $N_T$-extent (2), the reconstruction becomes easier when $\delta G(\tau)\rightarrow 0$ and $N_\tau\rightarrow\infty$; the finite lattice volume (3), as the spectral support is constrained by lattice momenta; quark masses (4) and lattice cut-off effects (5). The last, point (5), can be a problem especially for heavy systems, like bottomonia, where the cut-off region in the spectral function can be too close to separate it cleanly in the result.
Ideally the lattice data should be extrapolated to the continuum, extrapolated to the infinite volume limit - or at least have large volumes, be at the correct quark masses, have very large time extents and have very small statistical errors. The calculations below fulfill these quality criteria to varying degrees. But a careful evaluation is important in all cases and their discussion goes beyond this contribution.

\subsection{ { Dilepton rate and electrical conductivity }   } 

In the case of dilepton rates and electrical conductivities quenched and dynamical QCD results are available from multiple groups, see \cite{Aarts:2020dda} and \cite{Kaczmarek:2022ffn} for recent overviews.
The most common methods used in the available analyses are $\chi^2$-fits with additional constraints, MEM and the BG method. The fit Ansatz typically includes a transport peak contribution plus a peak of the Breit-Wigner type for the $\rho$ meson and the asymptotic behavior estimated using thermal field theory methods. Here also non-positive spectral function combinations and extra sum rule constraints can be implemented \cite{Brandt:2012jc}. 
Typically a main difficulty is to distinguish the transport from the meson peak contribution, see e.g. \cite{Brandt:2015aqk} for a discussion. The MEM studies naturally encode the positivity constraint, such that certain spectral function differences are not available for study. The default models, i.e. priors, typically scan a variety of example spectral functions along the lines of the Ansatz functions used in the studies using fits, see e.g. \cite{Aarts:2020dda} and references therein.
Comparing all available studies a consistent picture is emerging. In particular we show the collected results for the electrical conductivity in Fig.~\ref{fig:comp-el} and an example of a dilepton rate in the rightmost panel of the same figure \cite{Ding:2016hua}.

\begin{figure}
\centering
\includegraphics[width=0.44\textwidth]{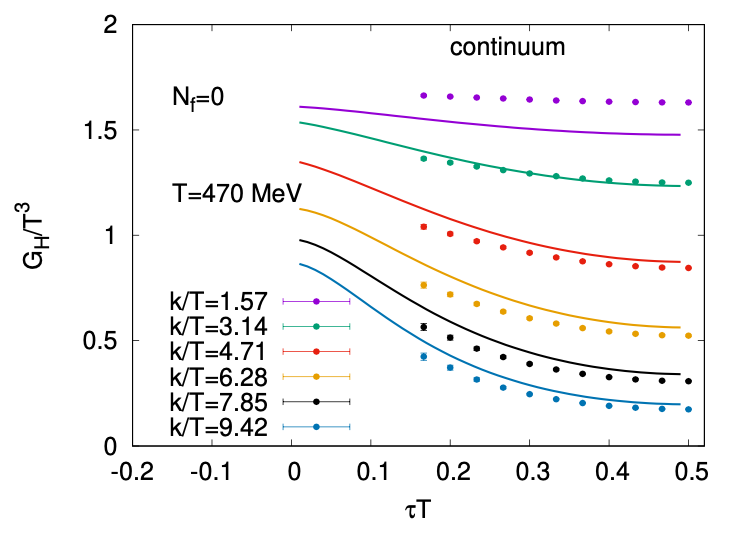}
\includegraphics[width=0.44\textwidth]{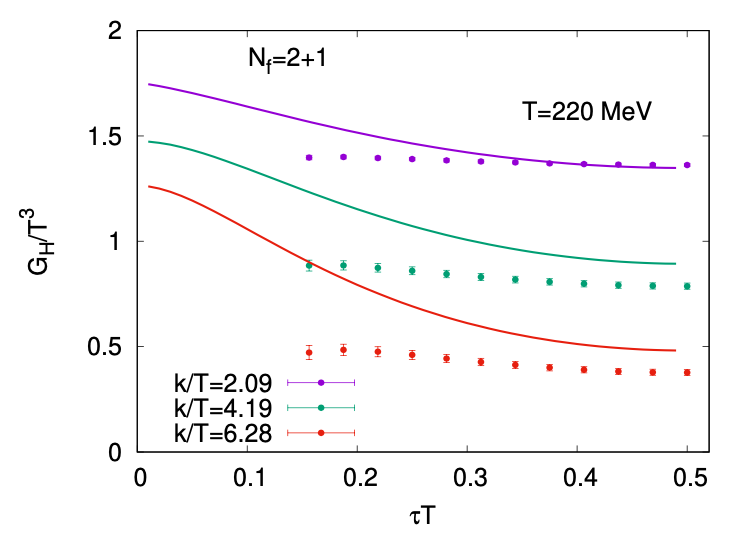}
\caption{Comparison of the quenched (left) and $n_f=2+1$ dynamical QCD (right) determinations of the improved estimator for the photon production rate with perturbative results \cite{Ali:2024xae}.}
\label{fig:phot-corr}
\end{figure}

\subsection{ {Photon production rate from the lattice}    } 

In the case of the photon rate the example shown in Fig.~\ref{fig:chi-landscape} illustrates that a better estimator that is more sensitive to the crucial $\omega=|\vec k|$ point is desirable. Such an estimator was worked out \cite{Ce:2020tmx},
 and goes back to the initial definition of Eq.~\ref{eq:photonrate}. To this extent, notice that at $T>0$, there are two independent components (longitudinal and transverse): 
\begin{equation}
\rho_L(\omega, k) \equiv\left(\hat{k}^i \hat{k}^j \rho^{i j}-\rho^{00}\right), \quad \textrm{and} \quad \rho_T(\omega, k) \equiv \frac{1}{2}\left(\delta^{i j}-\hat{k}^i \hat{k}^j\right) \rho^{i j}
\end{equation}
where $k \equiv|\boldsymbol{k}|$ and $\hat{k}^i=k^i / k$. Due to current conservation we have that $\omega^2 \rho^{00}(\omega, k)=k^i k^j \rho^{i j}(\omega, k)$ and therefore $\rho_L$ vanishes at light like kinematics, $\mathcal{K}^2=0$.
 This insight can be used to rewrite the photon rate with the simultaneous introduction of the new estimator:
\begin{equation}
\rho(\omega, k, \lambda)=2 \rho_T+\lambda \rho_L  \stackrel{\lambda=1}{=}-\rho^\mu{ }_\mu
\end{equation}
For any value of $\lambda$ we have
\begin{equation}
d \Gamma_\gamma(\boldsymbol{k})=\alpha \frac{d^3 k}{4 \pi^2 k} \frac{\rho(k, k, \lambda)}{e^{\beta k}-1}~.
\end{equation}
and our goal is to choose $\lambda$ such that the reconstruction becomes particularly easy. 
For example, by choosing $\lambda=2$ the spectral function we want becomes \cite{Ce:2020tmx,Ali:2024xae}
\begin{equation}
\rho_H(\omega, \vec{k})=2\left\{\rho_T(\omega, \vec{k})-\rho_L(\omega, \vec{k})\right\}
\end{equation}
which has the benefits that this estimator is $\rho_H(\omega, \vec{k})=0$ for $T=0$, due to the restoration of Lorentz symmetry, and only thermal effects contribute at $T>0$.  Further, asymptotically $\rho_H(\omega, \vec{k})\sim 1/\omega^4$ for $\omega\gg k$ ($\pi T$) and finally there is a possible extra sum rule constraint $
\int_0^{\infty} \mathrm{d} \omega \omega \rho_H(\omega)=0
$.
Furthermore, this estimator can be worked out in thermal perturbation theory at $\textrm{NLO + LPM}^{\textrm{LO}}$, see \cite{Jackson:2019yao} for details,
and therefore there is a perturbative benchmark to compare the spectral reconstruction of the non-perturbative lattice data to. 

\begin{figure}
\centering
\includegraphics[width=0.8\textwidth]{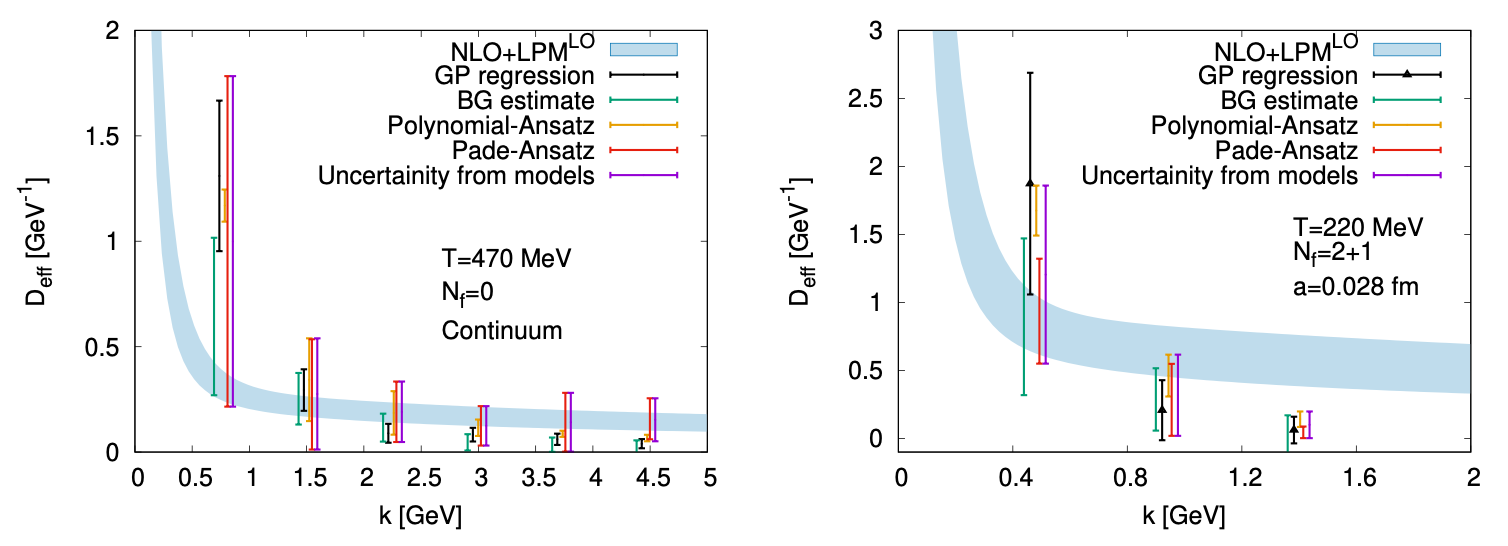}
\caption{Eﬀective diﬀusion coeﬃcient from all spectral reconstruction methods studied \cite{Ali:2024xae}, quenched (left) and $n_f=2+1$ dynamical QCD (right). For better visibility,
error bars at the same momentum value are slightly shifted on the horizontal axis.}
\label{fig:final}
\end{figure}

In our new calculation \cite{Ali:2024xae} we studied this estimator in quenched and dynamical QCD at finite temperature. The first set of results in $n_f=0$ quenched QCD uses ensembles at $a^{-1}= 9.4, 11.3, 14.1\,$GeV using Wilson-Clover fermions in the valence sector that were tuned to be close to $\kappa_c$ and with a temperature of $1.5~T_c$. The second set is in $n_f=2+1$ dynamical QCD on configurations with $a^{-1}= 7.04\,$GeV with Wilson-Clover quarks in the valence sector and HISQ quarks in the sea. The temperature in this case is $\simeq 1.2 T_{pc}$ and ${m_\pi=320\,}$MeV, $m_s=5\cdot m_\ell$. 
In the case of the quenched study the results are continuum extrapolated.
The results for the Euclidean correlator $G_H(\tau T)$ in both cases is shown in Fig.~\ref{fig:phot-corr} together with the mentioned perturbative result.
While in the quenched case the perturbative and lattice results match up rather well, especially at large spatial momenta, in the dynamical QCD case the differences are more pronounced. A further study after taking the continuum limit is desirable to clarify this issue.

In the next step, we employ all three basic strategies to perform the spectral reconstruction. In particular we perform $\chi^2$-fits (strategy II), Backus-Gilbert method reconstruction (strategy III) and Gaussian process regression (strategy I).
As highlighted before, the aim is to enable the study of the different systematics in all approaches with a maximal overview over the possible outcomes with the hope of arriving at a robust and conservative final result.
For the Gaussian process regression we perform a simultaneous reconstruction in ($\omega$, $k$) with continuity as the only constraint, so as to use the mildest possible constraints. For the $\chi^2$-fits we use two sets of Ansatz functions in terms of a polynomial fit to reproduce IR and UV results (based on the OPE) and a Pad\'e fit with the sum rule incorporated (based on OPE and AdS/CFT considerations), with the aim of including as much extra information as possible.
Finally, for the BGM reconstruction, we improve the reconstruction by rescaling the spectral function by the asymptotic behavior:
\begin{equation}
\frac{\rho_H^{\mathrm{BG}}(\omega, \vec{k})}{f(\omega, \vec{k})}=\sum_i q_i(\omega, \vec{k}) G_H\left(\tau_i, \vec{k}\right)
~~~\text{with rescaling fct.: }~~f(\omega, \vec{k})=\left(\frac{\omega_0}{\omega}\right)^4 \tanh \left(\frac{\omega}{\omega_0}\right)^5
\end{equation}

\begin{figure}
\centering
\includegraphics[width=0.6\textwidth]{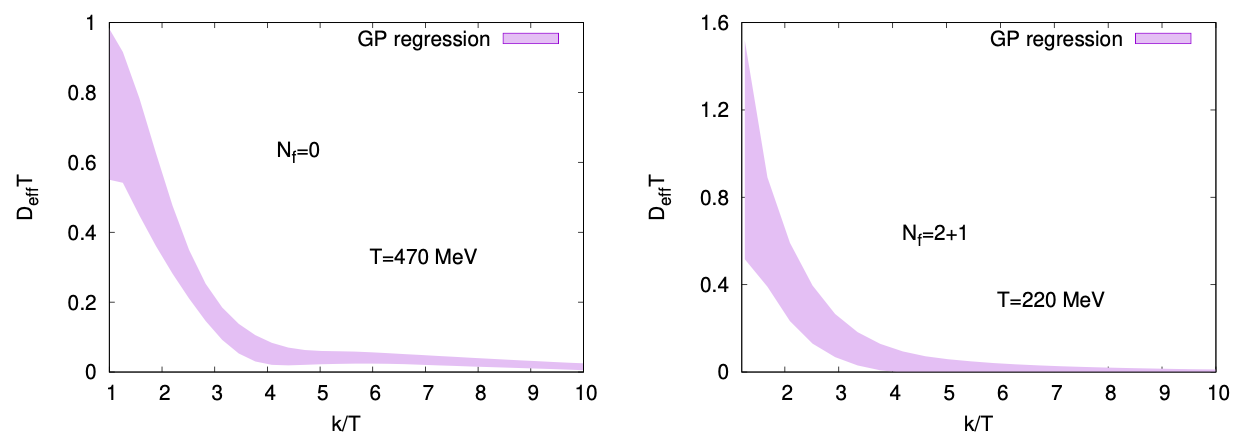}
\includegraphics[width=0.3\textwidth]{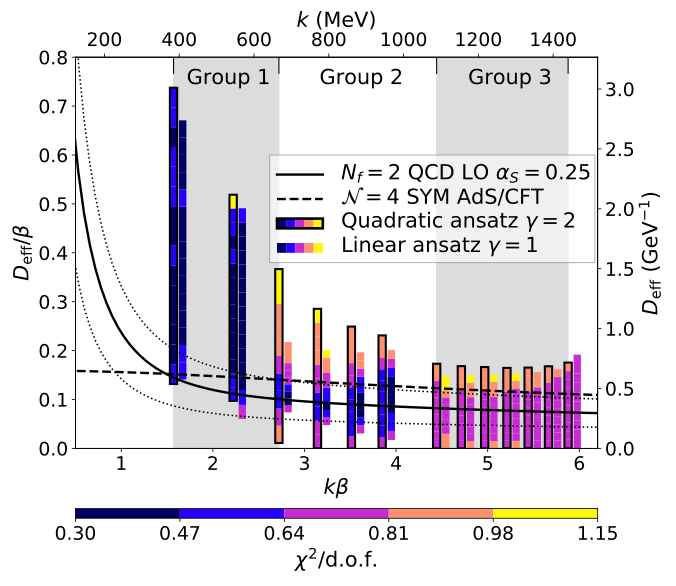}
\caption{The same as Fig.~\ref{fig:final} but in units $D_{\text{eff}}T=D_{\text{eff}}/\beta$. The left and middle panel are from \cite{Ali:2024xae} and the rightmost panel from \cite{Ce:2022fot}.}
\label{fig:phot-comp}
\end{figure}

The final results for the photon production rate obtained after analyzing the spectral functions reconstructed using all methods are collected in Fig.~\ref{fig:final}. Plotted is hereby the quantity:
\begin{equation}
D_{\mathrm{eff}}(k) \equiv \frac{\rho_H(\omega=k, k)}{2 \chi_q k}
\end{equation}
which is connected to the photon production rate via:
\begin{equation}
\frac{\mathrm{d} \Gamma_\gamma (\boldsymbol{k})}{\mathrm{d}^3 {k}}=  \frac{\alpha}{4 \pi^2 k} \frac{\rho(k, k, \lambda=2)}{e^{\beta k}-1}
=\frac{\alpha n_b(k) \chi_q}{\pi^2}(\sum_{i=1}^{N_f} Q_i^2)\, D_{\text {eff }}(k)~~.
\end{equation}
To compare these new and previous findings, results on $D_{\text{eff}}T=D_{\text{eff}}/\beta$ are shown in Fig.~\ref{fig:phot-comp}, which shows results in $n_f=2$ dynamical QCD \cite{Ce:2022fot}, see also \cite{Ce:2023oak} for the most recent update, in its rightmost panel. In general the comparison is favorable.

\section{Conclusions}

In this proceedings contribution we discussed the inverse problem to obtain (thermal) spectral functions from lattice correlators in general terms. Broadly highlighting applications and the connection to the dilepton and photon photon production rates in heavy-ion collisions, we sketched the general approach strategies and their challenges.
In the collected brief overview of the results on dilepton rates and the electrical conductivity we showed that a consistent picture between the different lattice calculations is emerging. One caveat here is that statistically speaking the data sets that went into all of these studies are not that significantly different, since the inverse problem is argued to be foremost a data problem, this consistent picture could still be challenged once order of magnitude(s) more precise data becomes available.
Finally we turned to recent developments on determining the photon production rate and highlighted an improved estimator that renders the inverse problem more doable. Focusing on our own studies in this direction our work on applying all basic strategies to the inverse problem
were presented and the new result shown, also in comparison to other existing results in dynamical QCD.
In the case of the photon rate the formulation of the improved estimator has made the study of the photon rate tractable. Similar success was achieved in the study of heavy quark diffusion \cite{Caron-Huot:2009ncn,Banerjee:2011ra,Francis:2015daa,Altenkort:2020fgs,Altenkort:2023oms}. In the future it is interesting to develop further such improved estimators. This goes in particular also to the use of smeared spectral function estimators.

\vspace{-2ex}
\section*{Acknowledgments}
This work is supported by the National Science and Technology Council (NSTC) of Taiwan under grant 113-2112-M-A49-018-MY3. Part of this contribution was gathered from presentations developed separately together with Maxwell T. Hansen, Alexander Rothkopf and Sinead Ryan. 
\vspace{-2ex}

\bibliographystyle{JHEP}

\bibliography{references}


\end{document}